\documentclass[11pt, onecolumn]{IEEEtran}
\usepackage[top=50mm, bottom=50mm, left=50mm, right=50mm]{geometry}
\usepackage{lineno}
\usepackage{amssymb}
\usepackage{amsmath}
\usepackage{amsthm}
\usepackage{epsfig}
\usepackage{graphicx}
\usepackage{graphics}
\usepackage{float}
\usepackage{multirow}
\usepackage{color}
\usepackage{lineno}
\usepackage{fullpage}
\usepackage[normalem]{ulem} 
\usepackage{makeidx}
\usepackage{xspace}
\usepackage{url}
\usepackage{wrapfig}
\usepackage{tabularx}
\usepackage{subcaption}
\usepackage{libertine}
\makeindex

\definecolor{darkred}{rgb}{1, 0.1, 0.3}
\definecolor{darkblue}{rgb}{0.1, 0.1, 1}
\definecolor{darkgreen}{rgb}{0,0.6,0.5}

\newcommand {\mm}[1] {\ifmmode{#1}\else{\mbox{\(#1\)}}\fi}

\newcommand{\cv}		{\mathrm{cv}}

\title{Analyzing Cross Validation In Compressed Sensing With Mixed Gaussian And Impulse Measurement Noise With L1 Errors}

\author{ Chinmay Gurjarpadhye$^{\sharp \ast}$, Shubhang Bhatnagar$^{\sharp \ast}$ and Ajit Rajwade$^\dag$\footnote{$\sharp$- CG and SB are first authors with equal contribution. AR acknowledges support from SERB Matrics Grant \#10013890.}}
\date{}

\begin{document}
\maketitle
\begin{abstract}
Compressed sensing (CS) involves sampling signals at rates less than their Nyquist rates and attempting to reconstruct them after sample acquisition. Most such algorithms have parameters, for example the regularization parameter in \textsc{Lasso}, which need to be chosen carefully for optimal performance. These parameters can be chosen based on assumptions on the noise level or signal sparsity, but this knowledge may often be unavailable. In such cases, cross validation (CV) can be used to choose these parameters in a purely data-driven fashion. Previous work analysing the use of CV in CS has been based on the $\ell_2$ cross-validation error with Gaussian measurement noise. But it is well known that the $\ell_2$ error is not robust to impulse noise and provides a poor estimate of the recovery error, failing to choose the best parameter. Here we propose using the $\ell_1$ CV error which provides substantial performance benefits given impulse measurement noise. Most importantly, we provide a detailed theoretical analysis and error bounds for the use of $\ell_1$ CV error in CS reconstruction. We show that with high probability, choosing the parameter that yields the minimum $\ell_1$ CV error is equivalent to choosing the minimum recovery error (which is not observable in practice). To our best knowledge, this is the first paper which theoretically analyzes $\ell_1$-based CV in CS.
\end{abstract}
%
\textbf{Key:} Compressed sensing, Cross validation, Impulse noise, $\ell_1$ error
%
\section{Introduction}
\label{sec:intro}
The goal of compressed sensing (CS) is to improve the efficiency of signal acquisition by enabling a signal to be reconstructed from a small number of its samples \cite{Candes2008}. It involves sampling the signal in a way such that most of the signal information is inherently available despite undersampling. The measured samples can be expressed in the form $\boldsymbol{y} = \boldsymbol{\Phi} \boldsymbol{x} + \boldsymbol{n}$, where $\boldsymbol{x} \in \mathbb{R}^N$ is a column vector representing the unknown signal, $\boldsymbol{y} \in \mathbb{R}^m$ is the vector of measurements, $\boldsymbol{n} \in \mathbb{R}^{m}$ is a noise vector such that $n_i \sim \mathcal{N}(0,\sigma^2_n/m)$, and $\boldsymbol{\Phi}$ is an $m \times N$ measurement matrix with $m \ll N$. The signal $\boldsymbol{x}$ is assumed to have a sparse representation in some $N \times N$ orthonormal basis $\boldsymbol{\Psi}$ so that $\boldsymbol{x} = \boldsymbol{\Psi \theta}$ where $\boldsymbol{\theta} \in \mathbb{R}^N$ is a sparse vector.

The signal $\boldsymbol{x}$ can be reconstructed from its measurements $\boldsymbol{y}$ using a variety of techniques such as \textsc{Lasso} \cite{THW2015} or greedy algorithms like Orthogonal Matching pursuit (OMP) \cite{OMP1} and its variants. \textsc{Lasso} seeks to minimize the cost function $\|\boldsymbol{y}-\boldsymbol{\Phi x}\|^2_2 + \lambda \|\boldsymbol{x}\|_1$ w.r.t. $\boldsymbol{x}$, where $\lambda$ is a regularization parameter. OMP seeks to minimize $\|\boldsymbol{x}\|_0$ s.t. $\|\boldsymbol{y}-\boldsymbol{\Phi x}\|_2 \leq \epsilon_n$ where $\epsilon_n$ is a regularization parameter dependent on the noise level. For optimal choice of the regularization parameter, many techniques rely on an estimate of the signal sparsity or noise level. However in real world scenarios, such information may not always be available. An alternative to this, proposed in \cite{Boufonos2007}, is using cross validation (CV), which is a purely data-driven technique. It proposes to set aside $m_{cv} < m$  measurements in $\boldsymbol{y}$ only for cross-validation (and not for reconstruction). These measurements are included in a $m_{cv} \times 1$ sub-vector $\boldsymbol{y_{cv}}$ with the corresponding $m_{cv} \times N$ sub-matrix $\boldsymbol{\Phi_{cv}}$. The remaining $m-m_{cv}$ measurements alone are used for compressive reconstruction, with each different value $\lambda$ of the parameter chosen from a candidate set $\Lambda$. For each parameter value $\lambda \in \Lambda$, the CV error $\epsilon_{cv,\ell_2,\lambda}:= \|\boldsymbol{y_{cv}}-\boldsymbol{\Phi_{cv}\hat{x}_{\lambda}}\|_2$ is computed where $\boldsymbol{\hat{x}_{\lambda}}$ is an estimate of $\boldsymbol{x}$ using parameter $\lambda$ in \textsc{Lasso}. The reconstruction $\boldsymbol{\hat{x}_{\lambda}}$ corresponding to the value $\lambda \in \Lambda$ which yielded the lowest value of $\epsilon_{cv,\ell_2,\lambda}$ is chosen as the final one. The work in \cite{Ward2009} proves theoretically using the Johnson-Lindenstrauss lemma \cite{Johnson1984} the close relationship between the data-driven CV error $\epsilon_{cv,\ell_2,\lambda}$ and the unobservable recovery error $\varepsilon_{x} := \|\boldsymbol{x}-\boldsymbol{\hat{x}_{\lambda}}\|_{2}$ for the case of zero measurement noise. A detailed analysis for the case of additive iid Gaussian noise in the measurements $\boldsymbol{y}$ has been presented in \cite{Zhang2014}, making use of the central limit theorem (CLT). The work in \cite{Zhang2014} calculates probabilistic bounds on the recovery error using the CV error, which theoretically justifies use of CV for CS in noisy signals. 

In this paper, we analyse the use of CV for signals with mixed heavy-tailed and Gaussian noise in the measurement vector $\boldsymbol{y}$. The problem is motivated by the fact that heavy-tailed noise is common in many compressive systems (see \cite{Tzagkarakis2019,Carrillo2016} and references therein). We observe that the $\ell_2$-based CV error $\epsilon_{cv,\ell_2,\lambda}$  fails to give an accurate estimate of the actual recovery error for such noise models and fails to pick the best value of regularization parameters. We propose using the $\ell_1$ based CV error $\epsilon_{cv,\ell_1,\lambda} := \|\boldsymbol{y_{cv}}-\boldsymbol{\Phi_{cv} \hat{x}_{\lambda}}\|_1$ instead, and demonstrate its significant performance benefits given impulse noise in the measurements. Most importantly, we derive the distribution of $\epsilon_{cv,\ell_1,\lambda}$ theoretically and establish its relationship to the recovery error $\|\boldsymbol{x}-\boldsymbol{\hat{x}_{\lambda}}\|_2$. As will be seen, our theoretical results match numerical simulation results very closely. 

\section{Problem Formulation and Motivation}
\label{sec:problem}
The main aim of this section is to motivate the need for $\ell_1$ CV errors. Throughout this work, we use a Gaussian sensing matrix, i.e $\forall (i,j), \Phi_{ij} \sim \mathcal{N}(0,1/m)$, though the analysis can be easily extended to other sub-Gaussian sensing matrices. As mentioned earlier, we consider a noise model which consists of a mixture of additive iid Gaussian noise from $\mathcal{N}(0,\sigma^2_n/m)$ and impulse noise. The latter is modeled as the product of a discrete random variable which takes on values in $\{-1,+1,0\}$ and a Gaussian random variable with a large mean as compared to $\sigma_n, \|\boldsymbol{x}\|_{\infty}$. Specifically, $\eta_i$ which is the $i^{th}$ element of the noise vector $\boldsymbol{\eta}$, is given as,
\begin{eqnarray}
\eta_i = n_{i} + B_{i}G_{i}, \\
B_i = 
\begin{cases}
    +1 & \mbox{w.p. }  b/2\\
    0 & \mbox{w.p. }  1-b\\
    -1 & \mbox{w.p. }  b/2,\\
\end{cases}
\label{eq:impulse_noise1}
\end{eqnarray}
where $n_{i} \sim \mathcal{N}(0,\sigma_{n}^2/m)$, $G_{i} \sim \mathcal{N}(\mu_{g},\sigma_{g}^2)$. 
Here, we choose the probability $b$ to be small, due to the sparse nature of impulse noise in many applications, and $\mu_{g}$ is chosen to be large compared to $\sigma_n,\sigma_g, \|\boldsymbol{x}\|_{\infty}$, since impulse noise often has very large magnitude as compared to the signal.  
\begin{figure}[h]
    \centering
    \includegraphics[width=9cm, height=6cm]{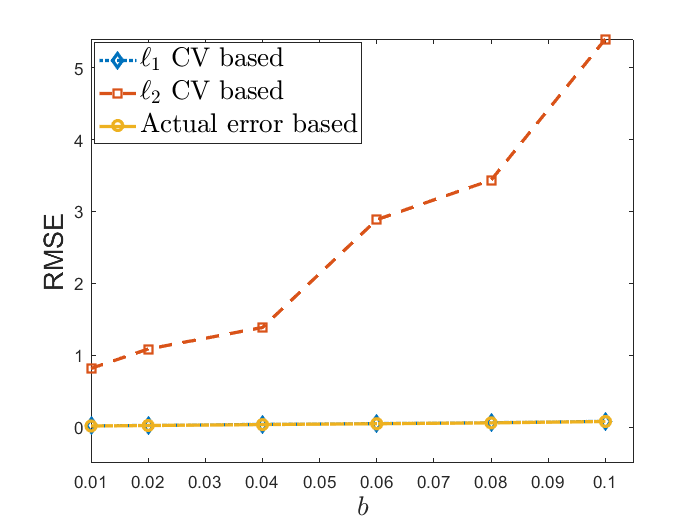}
    \caption{Using $\ell_1$-based CV error for parameter selection gives a much better RMSE than $\ell_2$-based CV error, in presence of different occurrence probabilities (i.e., $b$) of impulse noise. The plots for $\ell_1$-based CV error and true error nearly overlap.}
\label{fig:L1_L2_different_b}
\end{figure}
Next, we present simulation results for CS based recovery of a signal with different amounts of impulse noise in the measurements. We perform the reconstruction using the following version of the \textsc{Lasso} with an $\ell_1$-based data fidelity term:
\begin{equation}
\boldsymbol{\hat{x}_{\lambda}} = \textrm{argmin}_{\boldsymbol{x}} \|\boldsymbol{y}-\boldsymbol{\Phi x}\|_1 + \lambda \|\boldsymbol{x}\|_1,
\label{eq:L1_LASSO}
\end{equation}

in which we select the regularisation parameter $\lambda$ based on (1) minimum $\ell_1$ CV error, (2) minimum $\ell_2$ CV error, and (3) the minimum actual recovery error (implausible in real world, but useful for benchmarking). We use the parameters $\mu_{g}=700, \sigma_{n}=0.5, \sigma_{g}=100, m=420, N=1200, m_{cv} =20, s = \|\boldsymbol{x}\|_0 = 50,$ for the experiment. In Fig. 1, we conduct a comparison between the performance of $\ell_1$- and $\ell_2$-based CV for reconstruction using Eqn. \ref{eq:L1_LASSO} with different amounts of impulse noise in the signal, by varying $b \in \{0.01,0.02,...,0.09,0.1\}$. In this experiment, the non-zero elements of $\boldsymbol{x}$ were chosen iid from
$\mathcal{N}(0,10)$. 
\begin{figure}[h]
    \centering
    \includegraphics[width=9cm, height=6cm]{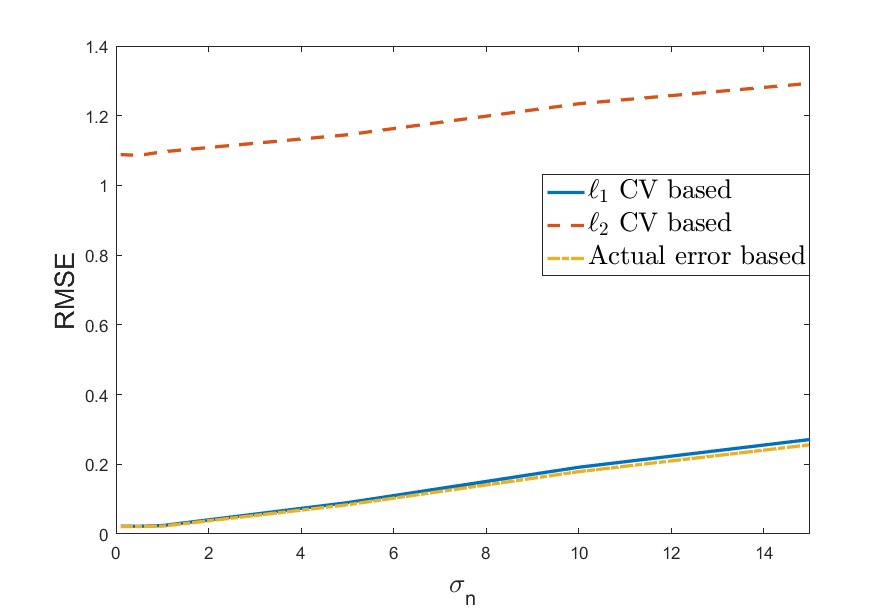}
    \caption{Using $\ell_1$-based CV error for parameter selection gives a much better RMSE than $\ell_2$-based CV error, in presence of varying amounts of Gaussian noise ($\sigma_{n}$) given a fixed, small occurrence of impulse noise ($b=0.02$).}
   \label{fig:L1L2True}
\end{figure}
The results in Fig. 1 show that the  parameter $\lambda$ chosen using $\ell_1$-based CV error $\epsilon_{cv,\ell_1,\lambda}$ gives a reconstruction whose RMSE almost coincides with the actual error for a large number of possible values of $b$. This  implies that $\ell_1$-based CV error is fairly robust to impulse noise and provides a significantly superior estimate of the real recovery error compared to $\ell_2$-based CV error.

Further, we compare the RMSE error for signal reconstruction using $\lambda$ chosen using  
$\ell_1$ based CV error and  $\ell_2$ based CV error for varying amount of non-impulse noise. For this experiment, we fix the probability of occurrence of impulse noise at  $b=0.02$ and vary $\sigma_n$ from 0 to 15. All the other parameters are same as the last experiment and non-zero elements of $\boldsymbol{x}$ are chosen iid from $\mathcal{N}(0,100)$.
This figure further impresses the robustness of reconstruction using $\ell_1$-based CV error compared to reconstruction using $\ell_2$-based CV error. 

\section{Theoretical Results}
Having seen some experimental results which strongly motivate the use of $\ell_1$-CV error, we now analytically derive a relationship between $\epsilon_{cv,\ell_1,\lambda} := \|\boldsymbol{y_{cv}}-\boldsymbol{\Phi_{cv} \hat{x}_{\lambda}}\|_1$ and the recovery error $\varepsilon_x := \|\boldsymbol{x}-\boldsymbol{\hat{x}_{\lambda}}\|_2$. This relationship holds with high probability (via the CLT) for large values of $m_{cv}$. In Lemma 1, we derive the distribution of the $\epsilon_{cv,\ell_1,\lambda}$ in terms of $\varepsilon_x$ and validate this result with relevant experiments. In Theorem 1, we use this distribution to obtain a high-probability confidence interval on the recovery error. We also give some experimental results to validate and illustrate the significance of these bounds. The distribution of the difference between the cross validation errors of two signal estimates is derived in Lemma 2 using CLT. This Lemma is crucially used to derive the probabilistic result in Theorem 2 which theoretically backs the use of $\ell_1$-CV error to choose the optimal regularization parameter.

\textbf{Lemma 1:} Assuming that $\mu_{g} \gg \sigma_g, \sigma_n, \varepsilon_x $ and that $m_{cv}$ is sufficiently large, we have $\epsilon_{cv,\ell_1,\lambda} \sim \mathcal{N}(\mu, \sigma^{2})$, where \\
$\mu= bm_{cv}\mu_{g} + (1-b)m_{cv} \sqrt{\frac{2}{m \pi}(\varepsilon^2_{x}+\sigma^2_{n})}$, \\
$\sigma^2= m_{cv}\Big(\frac{1}{m}(1 - (1-b)^2\frac{2}{\pi} )(\varepsilon^{2}_{x}+\sigma^2_n) + b(\sigma^{2}_{g}+(1-b)\mu^{2}_{g}) 
\\-2b(1-b)\mu_{g}\sqrt{\frac{2}{m \pi}(\varepsilon^2_{x}+\sigma^2_{n})}\Big)$. $\blacksquare$ \\
\begin{figure}[h]
    \centering
    \includegraphics[width=9cm, height=6cm]{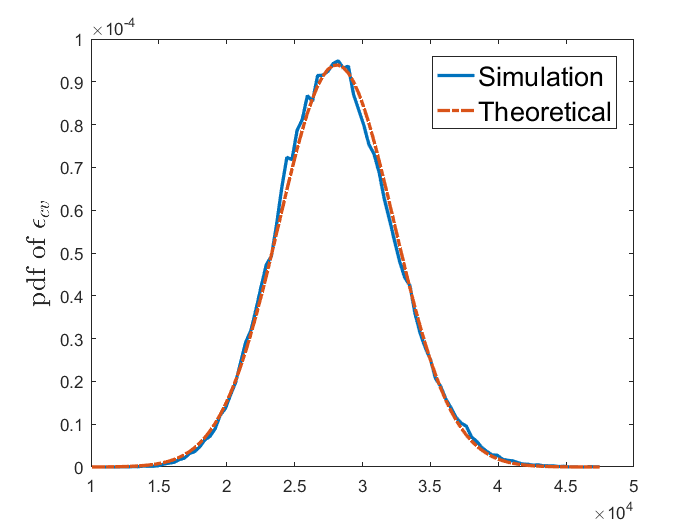}
    \caption{The simulated (obtained over $10^5$ noise realizations) and theoretically obtained pdf of $\epsilon_{cv,\ell_1,\lambda}$ (cf: Lemma 1) match closely.}
    \label{fig:lemma1_fig}
    \end{figure}
    \begin{figure}
     \centering
    \includegraphics[width=9cm, height=6cm]{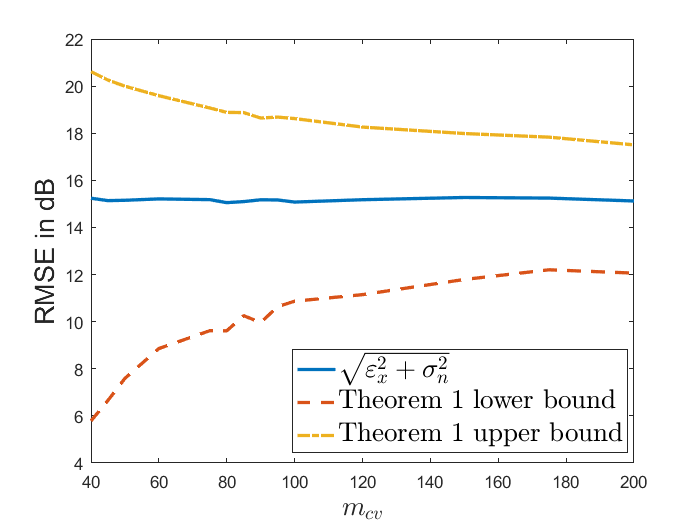}
    \caption{Empirical demonstration of confidence intervals from Theorem 1.}
    \label{fig:theorem1_fig}
\end{figure}
The various parameters in Lemma 1 have been earlier defined in Sec. \ref{sec:problem}. The proof of this lemma (in Sec. \ref{sec:proof}) uses many properties of the absolute value of the Gaussian distribution \cite{foldedND,Leone1961} followed by the Central limit theorem (CLT). Even if the CLT is an asymptotic result, we have observed excellent agreement between the empirically observed distribution of $\epsilon_{cv,\ell_1,\lambda}$ and the distribution predicted by Lemma 1, even at reasonable values of $m_{cv}$. This is seen in Fig. \ref{fig:lemma1_fig}, where we empirically compute the pdf of $\epsilon_{cv,\ell_1,\lambda}$ for 
$N=1200,m=800,m_{cv}=400$, given a single signal and $10^{5}$ noise realizations. This clearly corroborates the correctness of Lemma 1. Also, we observe that our assumption that $\mu_{g} \gg \varepsilon_{x}$ typically holds for a wide range of values of the regularization parameter( $\lambda \in \{0.001,0.01,...10000\}$) in our simulations. This is because impulses have very large magnitudes (as compared to the signal), and the reconstruction error is typically not of the same order as the impulse magnitude.

\begin{figure}
    \centering
    \includegraphics[width=9cm, height=6cm]{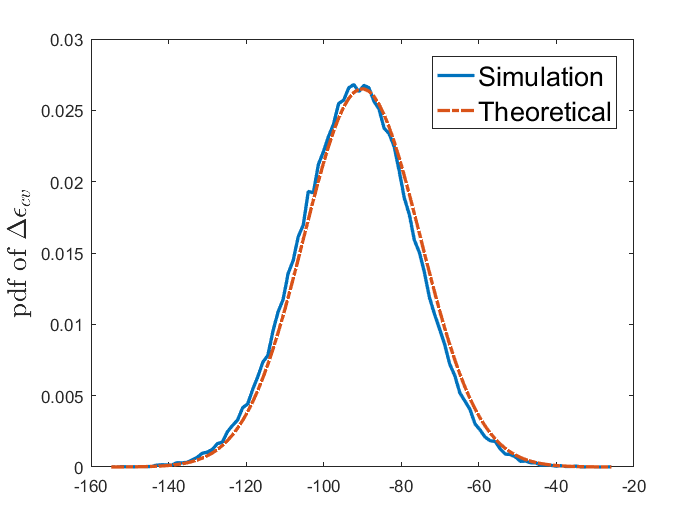}
    \caption{The simulated (obtained over $10^5$ noise realizations) and theoretically obtained pdf of $\Delta\epsilon_{cv} :=\epsilon_{\mathrm{cv}}^{p}-\epsilon_{cv}^{q}$ (cf: Lemma 2) match closely.}
    \label{fig:lemma2_fig}
\end{figure}

\begin{figure}
    \centering
    \includegraphics[width=9cm, height=6cm]{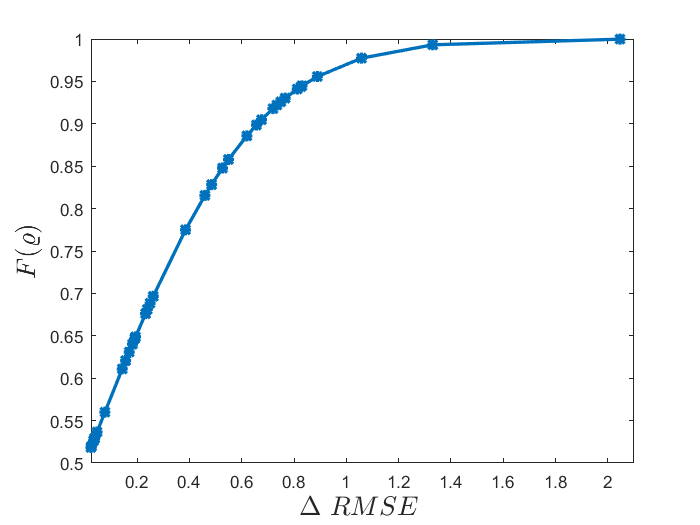}
    \caption{Probability that $\epsilon_{cv}^{p} \geq \epsilon_{cv}^{q}$ as given by Theorem 2, plotted against $\Delta RMSE := \frac{|\varepsilon_{x}^{p}-\varepsilon_{x}^{q}|}{\|\boldsymbol{x}\|_2}$.}
    \label{fig:theorem2_fig}
\end{figure}

\textbf{Theorem 1}: Assuming that $\mu_{g} \gg \sigma_g, \sigma_n, \varepsilon_x $, $m_{cv}$ is sufficiently large, and using $\epsilon_{cv}$ as shorthand for $\epsilon_{cv,\ell_1,\lambda}$, the following confidence interval holds with probability $\textrm{erf}(\varrho/\sqrt{2})$:
$ \frac{\sqrt{m}}{m_{cv}}\frac{\epsilon_{cv}- p(\varrho,+)}{h(\varrho,+)} \leq \sqrt{\varepsilon_{x}+\sigma_{n}^{2}} \leq \frac{\sqrt{m}}{m_{cv}} \frac{\epsilon_{cv} - p(\varrho,-)}{h(\varrho,-)}.$, where
\begin{align*}
    & p(\varrho,\pm):= m_{cv}b\mu_{G}  \pm \varrho \sqrt{m_{cv}b\left(\sigma_{G}^{2}+(1-b)\mu_{G}^{2}\right)}  \\
    & h(\varrho,\pm) := (1-b)\sqrt{\frac{2}{\pi}}\pm \varrho \sqrt{\frac{\left(1 - (1-b)^{2}\frac{2}{\pi}\right)}{m_{cv}}},
\end{align*}
Here, $\textrm{erf}(u) := \frac{1}{\sqrt{\pi}} \int_{-u}^{u} \mathrm{e}^{-t^{2}}dt$ denotes the error function and $\varrho$ is a free parameter. One can observe that choosing a higher value of $\varrho$ gives a looser bound but the bound holds with higher probability and vice versa for a lower value of $\varrho$. Furthermore, the width confidence interval from Theorem 1 denoted by $\mathcal{C}$ is given by \begin{align*}w := \frac{\sqrt{m}}{m_{cv}}\bigg(\frac{\epsilon_{cv}(h(\varrho,+) - h(\varrho,-))}{h(\varrho,-)h(\varrho,+)} +  \frac{p(\varrho,+)h(\varrho,-) - p(\varrho,-)h(\varrho,+) }{h(\varrho,-)h(\varrho,+)}\bigg)\end{align*}. 
As shown at the end of the proof of Theorem 1 in Sec. \ref{sec:proof}, this confidence interval drops to 0 in the limit as $m_{cv}$ tends to infinity. The proof of Theorem 1 crucially uses Lemma 1. In Fig. \ref{fig:theorem1_fig}, we demonstrate the upper and lower bounds as per Theorem 1 and the empirical recovery error for $N=1200, 40 \leq m_{cv} \leq 200, s=\|\boldsymbol{x}\|_0 = 50, b = 0.1, \sigma_n=0.5, \mu_g=700, \sigma_g=100, \varrho =3$. We note that similar results can be obtained for other parameters as well. For generating Fig. \ref{fig:theorem1_fig}, we have averaged over 1000 instances, with new realizations of all random variables in each instance. This figure as well as Theorem 1 both predict that the bounds become tighter with increase in $m_{cv}$, which is very intuitive.
\\
\textbf{Lemma 2:} Let $\hat{\boldsymbol{x}}^{p}$ and $\hat{\boldsymbol{x}}^{q}$ be two recovered signals with their respective $\ell_1$-CV errors $\epsilon^p_{cv},\epsilon^q_{cv}$ and respective true recovery errors $\varepsilon_p, \varepsilon_q$. Define $\Delta \boldsymbol{x}^p := \boldsymbol{x}-\boldsymbol{x}^p, \Delta \boldsymbol{x}^q := \boldsymbol{x}-\boldsymbol{x}^q$. Assuming that $\mu_{g} \gg \sigma_g, \sigma_n, \varepsilon_p,\varepsilon_q$ and that $m_{cv}$ is sufficiently large, we have
$\Delta \epsilon_{cv} := \epsilon_{\mathrm{cv}}^{p}-\epsilon_{cv}^{q} \sim \mathcal{N}\left(\mu, \sigma^{2}\right)$ where 
\begin{align*}
\mu= & (1-b){m_{cv}}K_{1}\left(\sigma_{p}-\sigma_{q}\right) \\ 
\sigma^{2}= & (1-b)m_{cv}\left( \sigma^{2}_{p} + \sigma^{2}_{q} -2\rho_1 \sigma_{p} \sigma_{q} \right)  +  m_{cv}\frac{b}{m} \left(\varepsilon^{2}_{p} + \varepsilon^{2}_{p} \right ) \\
& -\frac{2bm_{cv}}{m} \langle \Delta x^{p} , \Delta x^{q}\rangle  - m_{cv}((1-b) K_1 (\sigma_p - \sigma_q))^{2}\\
 \rho_1 = &\frac{\sigma_{p}\sigma_{q}}{\pi} \left(\pi \rho_2 -2\rho_2 \textrm{tan}^{-1}\left(\frac{\sqrt{1-\rho^{2}_2}}{\rho_{2}}\right) + 2\sqrt{1-\rho^{2}_2}\right) \\
  \rho_2 = & \frac{\sigma^{2}_n+\left\langle\Delta \mathbf{x}^{p}, \Delta \mathbf{x}^{q}\right\rangle}{m\sigma_p\sigma_q}\\
  \sigma_{p}=&\sqrt{\frac{\varepsilon_{p}^{2}+\sigma_{n}^{2}}{m}}, \qquad \sigma_{q}=\sqrt{\frac{\varepsilon_{q}^{2}+\sigma_{n}^{2}}{m}}, \qquad
K_{1} = \sqrt{\frac{2}{\pi}}. \blacksquare
\end{align*}
The proof of Lemma 2 can be found in Sec. \ref{sec:proof}. We demonstrate the result of Lemma 2 in Fig. \ref{fig:lemma2_fig} for $N=1200, m=440,
m_{cv}=40, b=0.1,
\sigma_n=0.5,
\sigma_g=100,
\mu_g=700.$
We conduct some experiments similar to the ones in Fig. \ref{fig:lemma1_fig} which demonstrate that the distribution of $\Delta \epsilon_{\mathrm{cv}}$ obtained in Lemma 2 indeed matches very closely with empirical results corroborating our assumptions.
\\
The following theorem shows that if the $\ell_1$ CV error of one recovered signal is larger than that of another recovered signal, then with high probability, the $\ell_2$ recovery errors for those signals follow the same order. 
\newline
\textbf{Theorem 2: }Let $\hat{\boldsymbol{x}}^{p}$ and $\hat{\boldsymbol{x}}^{q}$ be two recovered signals, with (unobservable) recovery errors $\varepsilon^p_x, \varepsilon^q_x$ and corresponding cross-validation errors $\epsilon^p_{cv},\epsilon^q_{cv}$. Assume $\mu_{g}>> \sigma_g, \sigma_n, \varepsilon_p,\varepsilon_q, $ and $m_{cv}$ is sufficiently large. If $\varepsilon_{x}^{p} \geq \varepsilon_{x}^{q}$, then it holds with probability $F(\varrho)$ that
$\epsilon_{cv}^{p} \geq \epsilon_{cv}^{q}$, where $\varrho =\frac{\mu}{\sigma}$
where, $\mu,\sigma$ are as defined in Lemma 2 and $F$ is the standard Gaussian CDF. $\blacksquare$
\\
From the expressions of $\mu,\sigma,\varrho$, one can observe that the confidence with which  \( \varepsilon_{\mathrm{x}}^{p} \geq \varepsilon_{\mathrm{x}}^{q} \) holds increases monotonically with \(m_{cv}\) and tends to 1 as \(m_{cv}\) increases. This is because, with all other parameters fixed, $\frac{\mu}{\sigma}$ is proportional to $\sqrt{m_{cv}}$. In Fig \ref{fig:theorem2_fig} we demonstrate this idea for
$N=1200, m=420, m_{cv}=20, \|\boldsymbol{x}\|_0 = 50,
b=0.05, \sigma_n=0.5, \mu_{g}=1000, \sigma_{g}=20$.
It is evident from Fig.\ref{fig:theorem2_fig} that, as the difference in the cross validation error of two recovered signals increases, the likelihood that their recovery error follows the same order increases. This observation is very crucial as it strongly supports the idea that $\ell_{1}$-CV error is a very good metric for choosing the optimal regularization parameter in our experiments.

\section{Proofs}
\label{sec:proof}
In this section, we present the proofs of Lemma 1, Theorem 1, Lemma 2 and Theorem 2 from the previous section. 
\\
\textbf{Assumption 1:} As discussed in the problem formulation, we assume the mean of the Gaussian noise to be large when compared to the values in signal $\boldsymbol{x}$ and to the Gaussian noise variance which translates to the following approximation,
\begin{align*}
    |\sum_{j=1}^{N}\   A_{\cv,i,j}\Delta x_{j} + n_{i} + G_{i}| \sim \sum_{j=1}^{N}\   A_{\cv,i,j}\Delta x_{j} + n_{i} + G_{i} \ \text{and} \\
   |\sum_{j=1}^{N}\   A_{\cv,i,j}\Delta x_{j} + n_{i} - G_{i}| \sim \sum_{j=1}^{N}\  - A_{\cv,i,j}\Delta x_{j} - n_{i} + G_{i}    
\end{align*}
\textbf{Remark 1:} Throughout these proofs, we come across the folded Gaussian distribution, i.e. the distribution of $|X|$ if $X \sim \mathcal{N}(\mu,\sigma^2)$. We note that that the mean, $\mu '$ of this distribution is given as $\mu ' =  \sigma\sqrt{\frac{2}{\pi}} e^{-\mu^2 / 2\sigma^2} - \mu \left(1-2 \Phi\left(\frac{\mu}{\sigma} \right) \right)$, where $\Phi(\cdot)$ is the cdf of the Gaussian distribution. This is proved in \cite{foldedND,Leone1961}.
\\
Also note that all the expectations in the proofs that follow are over the noise instances as well as the instances for the randomly generated sensing matrix $\boldsymbol{A}$.

\subsection{Proof of Lemma 1}
The L1 cross validation error $\epsilon_{\cv}$ (shorthand for $\epsilon_{cv,\ell_1,\lambda}$ in the main paper) is given as follows,
\begin{align*}
    \epsilon_{\cv}= & \left \| \boldsymbol{y}_{\cv} - \boldsymbol{\Phi_{\cv} \hat{x}_{\lambda}} \right \|_1 = \sum_{i=1}^{m_{\cv}} \left |(\sum_{j=1}^{N} A_{\cv,i,j}\Delta x_{j}) + n_i + B_{i}G_{i}  \right|,
    \end{align*}
    where $\boldsymbol{x}$ is the true, unknown signal, $\boldsymbol{\hat{x}_{\lambda}}$ is its estimate and $\Delta x_j := x_j - \hat{x}_{\lambda}(j)$.
    For simplicity of notation we define $r_i$ as follows,
    \begin{align*}
    r_{i}= \left |\sum_{j=1}^{N}\   A_{\cv,i,j}\Delta x_{j} + n_i + B_{i}G_{i}\right |
    \end{align*}
    Using this definition we have,
    \begin{align*}
    \epsilon_{\cv} = \sum_{i=1}^{m_{\cv}}r_{i}
    \end{align*}
    Now we compute the value of $E[r_{i}]$,
    \begin{align}
    E[r_{i}]=\frac{b}{2} E[r_{i} \mid B_{i}=-1 ] + (1-b) E[r_{i} \mid B_{i}= 0 ] + \frac{b}{2} E[r_{i} \mid B_{i}=1]. \label{eq:mean1}
    \end{align}
    We first compute the value of middle term in \eqref{eq:mean1}:
         \begin{align*}
            E[r_{i} \mid B_{i}=0 ] =
    & E\left[|\sum_{j=1}^{N}   A_{\cv,i,j}\Delta x_{j} + n_i |\right].
        \end{align*}
        Here observe that the random variable  $(\sum_{j=1}^{N}   A_{\cv,i,j}\Delta x_{j} + n_i)$ has mean 0 and its variance can be computed as follows,
        \begin{align*}
            E\left [(\sum_{j=1}^{N}   A_{\cv,i,j}\Delta x_{j} + n_i )^2\right] = \sum_{j=1}^{N} \Delta x_{j}^{2} E(A_{\cv,i,j}^{2}) + E(n_i^{2}) &=\frac{1}{m}\left(\sum_{i}^{N}\Delta x_{j}^{2} + \sigma_{n}^{2}\right).\\
        \end{align*}
        Using the formula for the mean of a folded Gaussian, we have:
        \begin{align*}
            E(r_{i} \mid B_{i}=0)=\sqrt{\frac{2 (\sum_{j=1}^{N}\Delta x_{j}^{2} + \sigma_n^{2})}{\pi m}}.
        \end{align*}
        Next, we calculate the conditional expectation for the other 2 terms which have $B=\pm 1$-
        \begin{align*}
            E(r_{i} \mid B_{i}=\pm1)= E(|\sum_{j=1}^{N}\   a_{\cv,i,j}\Delta x_{j} + n_i \pm G_{i}|).
        \end{align*}
        We now use Assumption 1 to approximate $|\sum_{j=1}^{N}\   A_{\cv,i,j}\Delta x_{j} + n_i + G_{i}| \sim \sum_{j=1}^{N}\   A_{\cv,i,j}\Delta x_{j} + n_{i} + G_{i}$ and $|\sum_{j=1}^{N}\   A_{\cv,i,j}\Delta x_{j} + n_{i} - G_{i}| \sim \sum_{j=1}^{N}\  - A_{\cv,i,j}\Delta x_{j} - n_{i} + G_{i}$.
        Now $\sum_{j=1}^{N}\   A_{\cv,i,j}\Delta x_{j} + n_{i} + G_{i}$ is a sum of independent Gaussian Random variables, and hence is a Gaussian with the following mean:
        \begin{align*}
            E(A_{\cv,i,j} \Delta x_{j}) +E(n_{i}) + E(G_{i}) = \mu_{G}.
        \end{align*}
        Similarly,$\sum_{j=1}^{N}\   -A_{\cv,i,j}\Delta x_{j} - n_{i} + G_{i}$ is also a sum of independent Gaussian Random variables, and hence is a Gaussian with mean
        \begin{align*}
E( A_{\cv,i,j} \Delta x_{j}) -E(n_{i}) + E(G_{i}) = \mu_{G}.
        \end{align*}
So, we have 
\begin{align*}
E(r_{i} \mid B_{i}=\pm 1)= \mu_{G}.
\end{align*}
        We take $\varepsilon_{x}^{2} := \sum_{j=1}^{N} \Delta x_{j}^{2}$ which is the actual recovery error squared given the signal $\boldsymbol{x}$.\\

Using these and substituting in \eqref{eq:mean1} we get the mean of $r_{i}$ as-
    \begin{align*}
        E(r_{i})= &\frac{b}{2} E(r_{i} \mid B_{i}=-1 ) + (1-b) E(r_{i} \mid B_{i}=-1 ) + \frac{b}{2} E(|r_{i}| \mid B_{i}=1 )\\
        &= b \mu_{G} + (1-b) \sqrt{\frac{2}{m \pi} \left( \varepsilon_{x}^{2}+\sigma_{n}^{2}\right)}.
    \end{align*}
Next, we calculate the variance of $r_{i}$
\begin{align}
    Var(r_{i})=E(r_{i}^{2}) - E(r_{i})^{2}.
    \label{eq:vari1}
\end{align}
We first calculate $E(r_{i}^{2})$:
\begin{flalign}
     E(r_{i}^{2}) =\sum_{j=1}^{N} E(A_{\cv,i,j}^{2}) \Delta x_{j}^{2} + E(N_{\cv,i}^{2}) + E(B_{i}^{2}G_{i}^{2}) +  2 E(B_{i}G_{i}N_{\cv,i}) + \\\nonumber 2\sum_{j=1}^{N}\left(E(A_{\cv,i,j}N_{\cv,i}) \Delta x_{j}\right) + 2\sum_{j=1}^{N}\left(E(A_{\cv,i,j}B_iG_i) \Delta x_{j}\right)\\
     \overset{(a)}{=}\frac{\varepsilon_{x}^{2}}{m} + \frac{\sigma_{n}^{2}}{m} + E(B_{i}^{2}) E(G_{i}^{2})+0+0+0\\
     =\frac{\varepsilon_{x}^{2}+\sigma_{n}^{2}}{m} + b(\sigma_{G}^{2}+\mu_{G}^{2}).
\end{flalign}
Here equality marked as `(a)' follows because $N_{\cv,i},B_{i},G_{i}$ and $A_{\cv,i,j}$ are all independent of each other and $N_{\cv,i},B_{i}$ and $A_{\cv,i,j}$ have 0 mean.\\
Now calculating the term $E(r_{i})^{2}$:
\begin{align*}
    E(r_{i})^{2}=&b^{2}\mu_{G}^{2}+ (1-b)^{2}\frac{2}{m\pi}(\sigma_{n}^{2}+\varepsilon_{x}^{2}) + 2b(1-b)\mu_{G}\sqrt{\frac{2}{m\pi}(\sigma_{n}^{2}+\varepsilon_{x}^{2})}
\end{align*}
Substituting these terms in equation \eqref{eq:vari1}, we obtain $\textrm{Var}(r_{i})$ as follows:
\begin{align*}
    \textrm{Var}(r_{i})=& \left(1 - (1-b)^{2}\frac{2}{\pi} \right) \frac{\varepsilon_{x}^{2}+\sigma_{n}^{2}}{m} +  (b)(\sigma_{G}^{2}+(1-b)\mu_{G}^{2}) - 2b(1-b)\mu_{G}\sqrt{\frac{2}{m\pi}(\sigma_{n}^{2}+\varepsilon_{x}^{2})}.
\end{align*}
Now seeing that $\epsilon_{\cv}=\sum_{i=1}^{m_{\cv}}|r_{i}|$, and assuming that $m_{\cv}$ is large, we apply the Central limit theorem (CLT) to say that $\epsilon_{\cv}$ follows a normal distribution with the following mean and variance-
\begin{align*}
    \epsilon_{cv} \sim N(m_{\cv}E(r_{i}),m_{\cv}Var(r_{i}))
\end{align*}
with -
\begin{align*}
    E(r_{i}) &=b \mu_{G} + (1-b) \sqrt{\frac{2}{m \pi} \left( \varepsilon_{x}^{2}+\sigma_{n}^{2}\right)} \\
    \textrm{Var}(r_{i}) &= \left(1 - (1-b)^{2}\frac{2}{\pi}\right)\frac{\varepsilon_{x}^{2}+\sigma_{n}^{2}}{m} +  (b)(\sigma_{G}^{2}+(1-b)\mu_{G}^{2}) - 2b(1-b)\mu_{G}\sqrt{\frac{2}{m\pi}(\sigma_{n}^{2}+\varepsilon_{x}^{2})}.
\end{align*}
This completes the proof of Lemma 1. $\blacksquare$

\subsection{Proof of Theorem 1}
Now assuming $\mu_{1} = E(\epsilon_{\cv})$ and the standard deviation of $\epsilon_{\cv}$ as $\sigma_{1}$, we can say that with probability $ \textrm{erf} (\frac{\varrho}{\sqrt{2}})$ the following inequality holds:
\begin{align}
    -\varrho \leq \frac{\epsilon_{\cv} - \mu_{1}}{\sigma_{1}} \leq \varrho.
    \label{eq:ineq1}
\end{align}
Consider $K_{1} :=\left(1 - (1-b)^{2}\frac{2}{\pi}\right), K_{2} := b\left(\sigma_{G}^{2}+(1-b)\mu_{G}^{2}\right)$.
Consider the expression for the variance of $\epsilon_{cv}$ in Lemma 1. Using the inequality $\sqrt{a}+\sqrt{b}\geq \sqrt{a+b}$ where $a, b > 0$, the following inequalities hold:
\begin{align}
 &\sqrt{\frac{m_{\cv}}{m}K_1(\varepsilon_{x}^{2}+\sigma_{n}^{2})}+\sqrt{m_{\cv}K_{2}} \geq \sqrt{\frac{m_{\cv}}{m}K_{1}(\varepsilon_{x}^{2}+\sigma_{n}^{2})+ m_{\cv}K_{2}} 
    \label{eq:ineq2}
\end{align}
This is because $K_{1},K_{2},m,m_{\cv}$ are all positive quantities.\\
Consider. $\sigma_{2}=\sqrt{\frac{m_{\cv}}{m}K_{1}\varepsilon_{x}^{2}+\sigma_{n}^{2}}+\sqrt{m_{\cv}K_{2}}$. From inequality \eqref{eq:ineq2} we can say that $\sigma_{2}\geq \sigma_{1}$\\

Now using inequality \eqref{eq:ineq1} and \eqref{eq:ineq2} we arrive at the following inequality:
\begin{align}
   \nonumber & -\varrho \leq \frac{\epsilon_{\cv} - \mu_{1}}{\sigma_{2}} \leq \varrho\\
   \nonumber & -\sigma_{2}\varrho \leq \epsilon_{\cv} - \mu_{1} \leq \sigma_{2}\varrho\\
     \nonumber & \mu_{1} -\sigma_{2}\varrho \leq \epsilon_{\cv} \leq \mu_{1} + \sigma_{2}\varrho\\
    \label{eq:ineq4}
\end{align}

Now looking at the LHS of the above inequality, we get-
\begin{align}
     \nonumber & -\varrho \sqrt{\frac{m_{\cv}K_{1}}{m}(\varepsilon_{x}^{2}+\sigma_{n}^{2})} -\varrho\sqrt{m_{\cv}K_{2}} +m_{\cv}b\mu_{G}+(1-b)m_{\cv}\sqrt{\frac{2}{m\pi}\varepsilon_{x}^{2}+\sigma_{n}^{2}}  \leq \epsilon_{\cv} \\
     \nonumber & \sqrt{\varepsilon_{x}^{2}+\sigma_{n}^{2}}\left((1-b)m_{\cv}\sqrt{\frac{2}{m\pi}} -\varrho \sqrt{\frac{m_{\cv}K_{1}}{m}}\right)  \leq \epsilon_{\cv} +\varrho\sqrt{m_{\cv}K_{2}}  - m_{\cv}b\mu_{G} \\
    & \sqrt{\varepsilon_{x}^{2}+\sigma_{n}^{2}} \leq \frac{\sqrt{m}}{m_{\cv}}\left(\frac{\epsilon_{\cv} +\varrho\sqrt{m_{\cv}K_{2}}  - m_{\cv}b\mu_{G}}{\left((1-b)\sqrt{\frac{2}{\pi}} -\varrho \sqrt{\frac{K_{1}}{m_{\cv}}}\right)} \right)
\end{align}

Now seeing RHS of the inequality \eqref{eq:ineq4}
\begin{align*}
    & \epsilon_{\cv} \leq \varrho \sqrt{\frac{m_{\cv}K_{1}}{m}(\varepsilon_{x}^{2}+\sigma_{n}^{2})} + \varrho \sqrt{m_{\cv}K_{2}} + m_{\cv}b\mu_{G}+ (1-b)m_{\cv}\sqrt{\frac{2}{m\pi}(\varepsilon_{x}^{2}+\sigma_{n}^{2})}\\
    & \epsilon_{\cv} - \varrho \sqrt{m_{\cv}K_{2}} - m_{\cv}b\mu_{G}  \leq \sqrt{\varepsilon_{x}^{2}+\sigma_{n}^{2}}\left(\varrho \sqrt{\frac{m_{\cv}K_{1}}{m}} + (1-b)m_{\cv}\sqrt{\frac{2}{m\pi}}\right) \\
    & \sqrt{\varepsilon_{x}^{2}+\sigma_{n}^{2}} \geq \frac{\sqrt{m}}{m_{\cv}} \left( \frac{\epsilon_{\cv}  - \varrho \sqrt{m_{\cv}K_{2}} - m_{\cv}b\mu_{G}}{\varrho \sqrt{\frac{K_{1}}{m_{\cv}}} + (1-b)\sqrt{\frac{2}{\pi}}}\right). 
\end{align*}
Defining
\begin{align*}
    & p(\varrho,\pm):= m_{\cv}b\mu_{G}  \pm \varrho \sqrt{m_{\cv}K_{2}}  \\
    & h(\varrho,\pm) := (1-b)\sqrt{\frac{2}{\pi}}\pm \varrho \sqrt{\frac{K_{1}}{m_{\cv}}},
\end{align*}
We arrive at the following confidence interval on $\varepsilon_{x}$ (with probability $\textrm{erf}(\varrho/\sqrt{2})$):
\begin{align}
    & \frac{\sqrt{m}}{m_{\cv}}\frac{\epsilon_{\cv}- p(\varrho,+)}{h(\varrho,+)} \leq \sqrt{\varepsilon_{x}+\sigma_{n}^{2}} \leq \frac{\sqrt{m}}{m_{\cv}} \frac{\epsilon_{\cv} - p(\varrho,-)}{h(\varrho,-)}.
\end{align}
This proves Theorem 1. $\blacksquare$\\
We observe that the confidence interval length tends to 0 as we increase the value of $m_{cv}$, as expected. We prove this claim in the following steps-
\begin{align*}
& \frac{\sqrt{m}}{m_{\cv}} \left(\frac{\epsilon_{\cv} - p(\varrho,-)}{h(\varrho,-)}-\right) \frac{\sqrt{m}}{m_{\cv}}\left(\frac{\epsilon_{\cv}- p(\varrho,+)}{h(\varrho,+)}\right) \\
& =\frac{\sqrt{m}}{m_{\cv}} \left(\frac{\epsilon_{\cv}(h(\varrho,+) - h(\varrho,+))  -h(\varrho,+)p(\varrho,-)+h(\varrho,-)p(\varrho,+)}{h(\varrho,-)h(\varrho,+)}\right) \\
& = \frac{\sqrt{m}}{m_{\cv}}\left(\frac{ 2\varrho \epsilon_{\cv} \sqrt{\frac{K_{1}}{m_{\cv}}} + 2(1-b)\sqrt{\frac{2}{\pi}}\varrho\sqrt{K_{2}m_{\cv}}-2\sqrt{\frac{K_{1}}{m_{\cv}}}b\mu_{G}\varrho m_{\cv}}{(1-b)^{2}\frac{2}{\pi}-\varrho^{2}\frac{K_{1}}{m_{\cv}}} \right)\\
    &= \sqrt{m}\left(\frac{ 2\varrho \epsilon_{\cv} \sqrt{\frac{K_{1}}{m_{\cv}^{3}}} + 2(1-b)\sqrt{\frac{2}{\pi}}\frac{\varrho}{\sqrt{m_{\cv}}}\sqrt{K_{2}}-2\sqrt{\frac{K_{1}}{m_{\cv}}}b\mu_{G}\varrho}{(1-b)^{2}\frac{2}{\pi}-\varrho^{2}\frac{K_{1}}{m_{\cv}}} \right) \\
    &\text{Taking limit } m_{\cv} \to \infty \text{ we get-}\\
    &= m \left(\frac{\lim_{m_{\cv} \to \infty}2\varrho \epsilon_{\cv} \sqrt{\frac{K_{1}}{m_{\cv}^{3}}} + \lim_{m_{\cv} \to \infty}2(1-b)\sqrt{\frac{2K_{2}}{\pi m_{\cv}}}\varrho - \lim_{m_{\cv} \to \infty} 2\sqrt{\frac{K_{1}}{m_{\cv}}}b\mu_{G}\varrho}{\lim_{m_{\cv} \to \infty}2(1-b)\sqrt{\frac{2K_{2}}{\pi m_{\cv}}}\frac{\varrho}{\sqrt{m_{\cv}}}\sqrt{K_{2}}-\lim_{m_{\cv} \to \infty}\varrho^{2}\frac{K_{1}}{m_{\cv}}}\right)\\
    &=m\left( \frac{0+0+0}{(1-b)^{2}\frac{2}{\pi}+0}\right)\\
    &=0
\end{align*}
Hence, we say that the length of our confidence interval tends to 0 as we increase the value of $m_{\cv}$. 
\subsection{Proof of Lemma 2}
Here, we wish to obtain the distribution of $\Delta \epsilon_{cv}$. We have:
\begin{align*}
    \Delta \epsilon_{\cv} &= \epsilon^{p}_{\cv} - \epsilon^{q}_{\cv} \\
    &= \Sigma_{i=1}^{m_{\cv}}\left(|\Sigma_{j=1}^{N}A_{\cv,ij} \Delta x^{p}_{j} + n_{cv,i} + B_{\cv,i}G_{\cv,i}| - |\Sigma_{j=1}^{N}A_{\cv,ij} \Delta x^{q}_{j} + n_{cv,i} +  B_{\cv,i}G_{\cv,i}| \right).
\end{align*}
For simplicity we define $r^{p}_{i}$, $r^{q}_{i}$ and $r_{i}$ as follows,
\begin{align*}
    r^{p}_{i} & := |\Sigma_{j=1}^{N}A_{\cv,ij} \Delta x^{p}_{j} + n_{cv,i} + B_{\cv,i} G_{\cv,i}| \qquad r^{q}_{i} := |\Sigma_{j=1}^{N}A_{\cv,ij} \Delta x^{q}_{j} + n_{cv,i} + B_{\cv,i} G_{\cv,i}| \\
    r_i & := r^{p}_{i} - r^{q}_{i}.
\end{align*}
We now compute $E[r_i]$. To that end, we compute $E[r^{p}_i]$.
\begin{align}
E[r^p_i] = (1-b)E[r^{p}_i|B_i = 0] + \frac{b}{2}E[r^{p}_i|B_i = 1]+ \frac{b}{2}E[r^{p}_i|B_i = -1]. \label{eq:split1}
\end{align}
We expand each of the above terms. We first compute $E[r^{p}_i|B_i = 1]$ as follows,
\begin{align*}
   E[r^{p}_i|B_i = 1] &= E[|\Sigma_{j=1}^{N}A_{\cv,ij} \Delta x^{p}_{j} + n_{cv,i} + B_{\cv,i} G_{\cv,i}| |B_i = 1]\\
   &= E[|\Sigma_{j=1}^{N}A_{\cv,ij} \Delta x^{p}_{j} + n_{cv,i} + G_{\cv,i}|]\\
   &\overset{(a)}{=} E[\Sigma_{j=1}^{N}A_{\cv,ij} \Delta x^{p}_{j} + n_{cv,i} + G_{\cv,i}]\\
   &= \mu_g.
\end{align*}
Here (a) follows from Assumption 1. Similarly, one can get $E[r^{p}_i|B_i = -1] = \mu_g$. Now, we compute the first term in \eqref{eq:split1}.
\begin{align*}
   E[r^{p}_i|B_i = 0] &= E[|\Sigma_{j=1}^{N}A_{\cv,ij} \Delta x^{p}_{j} + n_{cv,i} + B_{\cv,i} G_{\cv,i}| B_i = 0] \\
   &= E[|\Sigma_{j=1}^{N}A_{\cv,ij} \Delta x^{p}_{j} + n_{cv,i}| ] \\
   &= E[|\Sigma_{j=1}^{N}A_{\cv,ij} \Delta x^{p}_{j} + n_{cv,i}| ] .
\end{align*}
Observe that $\Sigma_{j=1}^{N}A_{\cv,ij} \Delta x^{p}_{j} + n_{cv,i}$ is a Gaussian with mean 0 and variance $ \frac{\varepsilon^{2}_{p} + \sigma^{2}_n}{m}$, where $\varepsilon^{2}_{p}=\sum_{j=1}^{N}\Delta x_{j}^{2}$. Thus the absolute value is a folded Gaussian and using Remark 1 we get,
\begin{align*}
   E[r^{p}_i|B_i = 0] &= \sqrt{\frac{2}{\pi}} \sqrt{\frac{\varepsilon^{2}_{p} + \sigma^{2}_n}{m}}.
\end{align*}
For simplicity, here onward we use $K_1$ and $\sigma_p$ to denote $ \sqrt{\frac{2}{\pi}}$ and $\sqrt{\frac{\varepsilon^{2}_{p} + \sigma^{2}_n}{m}}$ respectively. Similarly, we denote $\sqrt{\frac{\varepsilon^{2}_{q} + \sigma^{2}_n}{m}}$ by $\sigma_q$. Thus, continuing from \eqref{eq:split1} we get,
\begin{align}
E[r^p_i] &= (1-b)E[r^{p}_i|B_i = 0] + \frac{b}{2}E[r^{p}_i|B_i = 1]+ \frac{b}{2}E[r^{p}_i|B_i = -1] \nonumber \\
& = (1-b) \sqrt{\frac{2}{\pi}} \sqrt{\frac{\varepsilon^{2}_{p} + \sigma^{2}_n}{m}} + \mu_g - \mu_g \nonumber \\
& = (1-b) K_1 \sigma_p. \nonumber 
\end{align}
Due to the symmetry we will have,
\begin{align}
E[r^q_i] & = (1-b) K_1 \sigma_q \nonumber
\end{align}
Thus from the definition of $r_i$ we get,
\begin{align}
E[r_i] & = (1-b) K_1 (\sigma_p - \sigma_q) \label{eq:meanri}
\end{align}
Having computed $E[r_i]$ we now compute $E[r^2_i]$ as follows,
\begin{align} 
E[r^2_i] = (1-b)E[r^{2}_i|B_i = 0] + \frac{b}{2}E[r^{2}_i|B_i = 1]+ \frac{b}{2}E[r^{2}_i|B_i = -1] \label{eq:split2}
\end{align}
We expand each of the above terms. We first compute $E[r^{2}_i|B_i = 1]$ as follows,
\begin{align}
&E[r^{2}_i|B_i = 1] \nonumber \\
& = E\left[\left(|\Sigma_{j=1}^{N}A_{\cv,ij} \Delta x^{p}_{j} + n_{cv,i} + B_{\cv,i} G_{\cv,i}| -  |\Sigma_{j=1}^{N}A_{\cv,ij} \Delta x^{q}_{j} + n_{cv,i} + B_{\cv,i} G_{\cv,i}|\right)^{2}|B_i = 1 \right] \nonumber \\
& = E\left[\left(|\Sigma_{j=1}^{N}A_{\cv,ij} \Delta x^{p}_{j} + n_{cv,i} + G_{\cv,i}| -  |\Sigma_{j=1}^{N}A_{\cv,ij} \Delta x^{q}_{j} + n_{cv,i} + G_{\cv,i}|\right)^{2} \right] \nonumber \\
& \overset{(a)}{=} E\left[\left(\Sigma_{j=1}^{N}A_{\cv,ij} (\Delta x^{p}_{j} - \Delta x^{q}_{j}) \right)^{2} \right] \nonumber \\
& = \frac{1}{m} \left(\varepsilon^{2}_{p} + \varepsilon^{2}_{p} -2 \langle \Delta \mathbf{x}^{p} , \Delta \mathbf{x}^{q}\rangle \right) \label{eq:eqpause1}
\end{align}
Here, (a) follows from assumption 1. Similarly we get,
\begin{align}
E[r^{2}_i|B_i = -1]  =  \frac{1}{m} \left(\varepsilon^{2}_{p} + \varepsilon^{2}_{q} -2 \langle \Delta \mathbf{x}^{p} , \Delta \mathbf{x}^{q}\rangle \right) \label{eq:eqpause2}
\end{align}
Now we compute the first term in \eqref{eq:split2} as follows,
\begin{align}
& E[r^{2}_i|B_i = 0] \nonumber \\
& =  E\bigg[ |\Sigma_{j=1}^{N}A_{\cv,ij} \Delta x^{p}_{j} + n_{cv,i}|^{2} + |\Sigma_{j=1}^{N}A_{\cv,ij} \Delta x^{q}_{j} + n_{cv,i}|^{2} \nonumber \\   
& \quad - 2|\Sigma_{j=1}^{N}A_{\cv,ij} \Delta x^{p}_{j} +  n_{cv,i}| |\Sigma_{j=1}^{N}A_{\cv,ij} \Delta x^{q}_{j} + n_{cv,i}| \bigg] \label{eq:eqpause3}
\end{align}
Recall that $\Sigma_{j=1}^{N}A_{\cv,ij} \Delta x^{p}_{j} + n_{cv,i} \sim \mathcal{N}(0,\sigma^{2}_{p})$ as argued earlier. Similarly we have $\Sigma_{j=1}^{N}A_{\cv,ij} \Delta x^{q}_{j} + n_{cv,i} \sim \mathcal{N}(0,\sigma^{2}_{q})$. So we now compute the remaining term in \eqref{eq:eqpause3}. For simplicity we define Gaussian random variables $X$ and $Y$ as follows,
\[
X = \Sigma_{j=1}^{N}A_{\cv,ij} \Delta x^{p}_{j} +  n_{cv,i} \qquad Y = \Sigma_{j=1}^{N}A_{\cv,ij} \Delta x^{q}_{j} +  n_{cv,i}
\]
Note that $X$ and $Y$ are 0 mean but correlated random variables. We further define random variables $X'$ and $Y'$ as follows,
\[
X' = X/\sigma_p \qquad Y' = Y/\sigma_q
\]
Notice that $X'$ and $Y'$ are normally distributed. Let $\rho $ be the covariance of $X'$ and $Y'$. We compute the value of $\rho$ later. We can write $X'$ and $Y'$ as follows,
\[
X' = U \qquad Y' = \rho U + \sqrt{1-\rho^{2}} V
\]
where $U$ and $V$ are independent and identically distributed Gaussian random variables, with mean 0 and variance 1. Observe that $E[|XY|] = \sigma_p \sigma_q E[|X'Y'|]$. We now compute $E[|X'Y'|]$ as follows,
\allowdisplaybreaks
\begin{align*}
&E[|X'Y'|] \\
&= E[|U(\rho U + \sqrt{1-\rho^2} V)|] \\
&= \sqrt{1-\rho^2} E[|U(K_3 U + V)|]  \ \ \ \ \left(\text{   Here } K_3 := \frac{\rho}{\sqrt{1-\rho^{2}}}\right)\\
&= \sqrt{1-\rho^2} \int_{-\infty}^{\infty} \int_{-\infty}^{\infty} |K_3 x^2 +xy| f_{U}(x) f_{V}(y) dy dx \\
&= \sqrt{1-\rho^2} \int_{-\infty}^{\infty} f_{U}(x) |x| \left(\int_{-\infty}^{\infty} |K_3 x +y|  f_{V}(y) dy\right) dx \\
& \overset{(a)}{=} \sqrt{1-\rho^2} \int_{-\infty}^{\infty} f_{U}(x) |x| \left(K_1 e^{\frac{-x^2 K^2_3}{2}} + K_3 x \left(\textrm{erf}\left( \frac{K_{3}x}{\sqrt{2}}\right)\right)\right) dx \\
&= \sqrt{1-\rho^2} \int_{-\infty}^{\infty} f_{U}(x) |x| K_1 e^{\frac{-x^2 K^2_2}{2}} + \int_{-\infty}^{\infty} f_{U}(x) |x| K_3 x \left( \textrm{erf}\left( \frac{K_{3}x}{\sqrt{2}}\right)\right) dx \\
&= \sqrt{1-\rho^2} \int_{-\infty}^{\infty} \frac{e^{-\frac{x^2}{2}}}{\sqrt{2\pi}} |x| K_1 e^{\frac{-x^2 K^2_2}{2}} + \int_{0}^{\infty} 2f_{U}(x) |x| K_3 x \left( \textrm{erf}\left( \frac{K_{3}x}{\sqrt{2}}\right)\right) dx\\ 
& \overset{(b)}{=} \sqrt{1-\rho^2} \left( \int_{-\infty}^{\infty} \frac{1}{\sqrt{2\pi}} |x| K_1 e^{\frac{-x^2 (K^2_3 + 1)}{2}} + \frac{2K_3}{\sqrt{2\pi}} \left[\frac{\sqrt{\pi}}{4} \times 2\sqrt{2} - \frac{1}{2\sqrt{\pi}}\left(2\sqrt{2} \textrm{tan}^{-1}\left(\frac{1}{K_3}\right) - \frac{(K_3 / \sqrt{2})}{\frac{1}{2}(1/2 + K^2_3/2)}\right) \right] \right)\\ 
&= \sqrt{1-\rho^2} \left (\sqrt{\frac{2}{\pi}} \sqrt{\frac{2}{\pi}} \times \frac{1}{K^2_3 + 1} + K_3 \left[1 - \frac{2}{\pi} \textrm{tan}^{-1}\left(\frac{1}{K_3}\right) + \frac{2}{\pi} \frac{K_3}{K^2_3 + 1}\right] \right) \\
&= \sqrt{1-\rho^2} \left ( \frac{2}{\pi} \frac{1}{K^2_3 + 1} + \frac{\rho}{\sqrt{1-\rho^2}} \left[1 - \frac{2}{\pi} \textrm{tan}^{-1}\left(\frac{\sqrt{1-\rho^2}}{\rho}\right) + \frac{2\rho}{\pi}\sqrt{1-\rho^2} \right]\right) \\
&= {\rho} - \frac{2\rho \textrm{tan}^{-1}(\sqrt{1-\rho^2}/\rho)}{\pi} + \frac{2\sqrt{1-\rho^2}}{\pi}. 
\end{align*}
Here, (a) follows  from Remark 1; (b) follows from \cite{Leone1961}. Now we compute the value of $\rho$ as follows:
\begin{align}
\rho = E[X'Y'] & = \frac{E[XY]}{\sigma^p \sigma^q} \nonumber \\
& = \frac{E[N_{\cv,ij}^2 + \Sigma_{j=1}^{N} A^{2}_{\cv,ij} \Delta x^{p}_{j} \Delta x^{q}_{j}]}{\sigma^p \sigma^q} \nonumber \\
& = \frac{1}{\sigma^p \sigma^q}\left(\frac{\sigma^{2}_{N}}{m} + \frac{\langle \Delta x^{p}_{j} \Delta x^{q}_{j} \rangle }{m}\right) \label{eq:rho} 
\end{align}
Now, substituting all terms in \eqref{eq:split2} we get,
\begin{align*}
E[r^{2}_{i}] = (1-b)\left( \sigma^{2}_{p} + \sigma^{2}_{q} -2\rho_1 \sigma_{p} \sigma_{q} \right) +  \frac{b}{m} \left(\varepsilon^{2}_{p} + \varepsilon^{2}_{p} -2 \langle \Delta x^{p}_{j} , \Delta x^{q}_{j}\rangle \right)
\end{align*}
where 
\[
\rho_1 = {\rho} - \frac{2 \rho \textrm{tan}^{-1}(\sqrt{1-\rho^2}/\rho)}{\pi} + \frac{2\sqrt{1-\rho^2}}{\pi} 
\]
Here, $\rho$ is as given in \eqref{eq:rho}. Also $mean(r_i) = E[r_i]$ is known from \eqref{eq:meanri}.
Thus we have the variance of $r_i$ as,
\begin{align*}
Var(r_i) & = E[r^{2}_i] - (E[r_i])^2 \\
& = (1-b)\left( \sigma^{2}_{p} + \sigma^{2}_{q} -2\rho_1 \sigma_{p} \sigma_{q} \right) +  \frac{b}{m} \left(\varepsilon^{2}_{p} + \varepsilon^{2}_{p} -2 \langle \Delta x^{p}_{j} , \Delta x^{q}_{j}\rangle \right) - ((1-b) K_1 (\sigma_p - \sigma_q))^{2} \\
\end{align*}
Having obtained the mean and variance of $r_i$, we now use the Central Limit Theorem to compute the distribution of $\Delta \epsilon_{\cv}$. 
\[
\Delta \epsilon_{\cv} \sim \mathcal{N}(\mu, \sigma^2) \sim \mathcal{N}(m_{\cv} mean(r_i), m_{\cv} Var(r_i))
\]
This completes the proof of Lemma 2. $\blacksquare$

\subsection{Proof of Theorem 2}
Theorem 2 follows directly from Lemma 2. 
\begin{align}
    Pr(\epsilon^{p}_{cv}>\epsilon^{q}_{cv}) & = Pr(\Delta \epsilon_{\cv} > 0) \nonumber \\
    & = \int^{\infty}_{0} \frac{1}{\sigma \sqrt{2\pi}} e^{-\frac{(x-\mu)^2}{2\sigma^2}} dx \nonumber
\end{align}
Here, $\mu$ and $ \sigma$ are as given in Lemma 2. Now substituting $x = (t\sigma+\mu) $, gives
\begin{align}
    Pr(\epsilon^{p}_{cv}>\epsilon^{q}_{cv}) & = \int^{\infty}_{-\mu/ \sigma} \frac{1}{ \sqrt{2\pi}} e^{-\frac{t^2}{2}} dx \nonumber \\
    & = \int^{\mu/ \sigma}_{-\infty} \frac{1}{ \sqrt{2\pi}} e^{-\frac{t^2}{2}} dx \nonumber \\
    & = \Phi(\frac{\mu}{\sigma})
\end{align}
This completes the proof of Theorem 2. $\blacksquare$
\section{Conclusion}
We provide a detailed theoretical and empirical analysis of the $\ell_{1}$-CV error when used for compressive reconstruction in presence of mixed impulse and Gaussian noise. Under some assumptions, we prove that the $\ell_{1}$ CV error follows a Gaussian distribution and further provide upper and lower bounds on the recovery error of a signal estimate in terms of the $\ell_{1}$-CV error and perform  simulations for generic parameters to validate these results. We justify the use of $\ell_{1}$-CV error for selecting the optimal parameter in signal reconstruction algorithms like $\ell_1$-\textsc{Lasso}, by proving that with high probability, the ordering of the actual recovery error of any two signal estimates is the same as the ordering of their $\ell_1$-CV errors.
\bibliographystyle{IEEE}
\bibliography{bibliography}
\end{document}